\documentclass{PoS}

\newcommand{\url}[1]{\href{#1}{#1}}

\newcommand{\gammapyversion}{v0.3}

\newcommand{\hess}{H.E.S.S.~}

\newcommand{\gammaray}{$\gamma$-ray~}

\newcommand{\gammapy}{\textit{Gammapy~}}

\usepackage{lineno}
\usepackage[procnames]{listings}
\usepackage{color}
\usepackage{xcolor}
\usepackage{enumitem}
\definecolor{keywords}{HTML}{007021}
\definecolor{comments}{RGB}{0,0,113}
\definecolor{strings}{HTML}{4070A1}
\definecolor{package}{HTML}{0C85B5}
\definecolor{lightgrey}{rgb}{0.98,0.98,0.98}
\definecolor{numbers}{HTML}{A30000}

\title{\gammapy -- A Python package for \gammaray astronomy}

\ShortTitle{\gammapy -- A Python package for \gammaray astronomy}

\author{
\speaker{Axel Donath}$^a$,
Christoph Deil$^a$,
Manuel Paz Arribas$^e$,
Johannes King$^a$,
Ellis Owen$^b$,
R\'egis Terrier$^c$,
Ignasi Reichardt$^c$ $^d$,
Jon Harris,
Rolf B\"uhler$^f$,
Stefan Klepser$^f$
\\
\llap{$^a$}MPIK, Heidelberg, Germany\\
\llap{$^b$}UCL-MSSL, Dorking, United Kingdom\\
\llap{$^c$}APC, University of Paris 7, France\\
\llap{$^d$}INFN, Padova, Italy\\
\llap{$^e$}Humboldt University, Berlin, Germany\\
\llap{$^f$}DESY, Zeuthen, Germany\\
E-mail:  \email{Axel.Donath@mpi-hd.mpg.de}, \email{Christoph.Deil@mpi-hd.mpg.de}
}

\abstract{

In the past decade imaging atmospheric Cherenkov telescope arrays such as
\href{http://www.mpi-hd.mpg.de/hfm/HESS/}{H.E.S.S.},
\href{https://magic.mpp.mpg.de/}{MAGIC},
\href{https://veritas.sao.arizona.edu/}{VERITAS}, as well as the
\href{http://fermi.gsfc.nasa.gov/}{\textit{Fermi}-LAT} space telescope have provided 
high quality images and spectra of the \gammaray universe.
Currently the \gammaray community is preparing to build the next-generation
Cherenkov Telecope Array (\href{https://www.cta-observatory.org/}{CTA}), which
will be operated as an open observatory.

\gammapy \gammapyversion \hspace{1pt} (available at \url{https://github.com/gammapy/gammapy} under the
open-source BSD license) is a new in-development Astropy affiliated package for
high-level analysis and simulation of astronomical \gammaray data. It is built
on the scientific Python stack (\href{http://www.numpy.org/}{Numpy},
\href{http://www.scipy.org/}{Scipy}, \href{http://matplotlib.org/}{matplotlib}
and \href{http://scikit-image.org/}{scikit-image}) and makes use of other
open-source astronomy packages such as \href{http://www.astropy.org/}{Astropy},
\href{http://cxc.cfa.harvard.edu/sherpa/}{Sherpa} and
\href{http://gammafit.readthedocs.org/en/latest/}{Naima} to provide a flexible
set of tools for \gammaray astronomers.

We present an overview of the \gammapy scope, development workflow, status,
structure, features, application examples and goals. We would like \gammapy to
become a community-developed project and a place of collaboration between
scientists interested in \gammaray astronomy with Python. Contributions welcome!

}

\FullConference{The 34th International Cosmic Ray Conference,\\
		30 July- 6 August, 2015\\
		The Hague, The Netherlands}

\begin{document}

\section{Introduction}
\begin{figure}
\centering
\includegraphics[height=0.7\textwidth, angle=270]{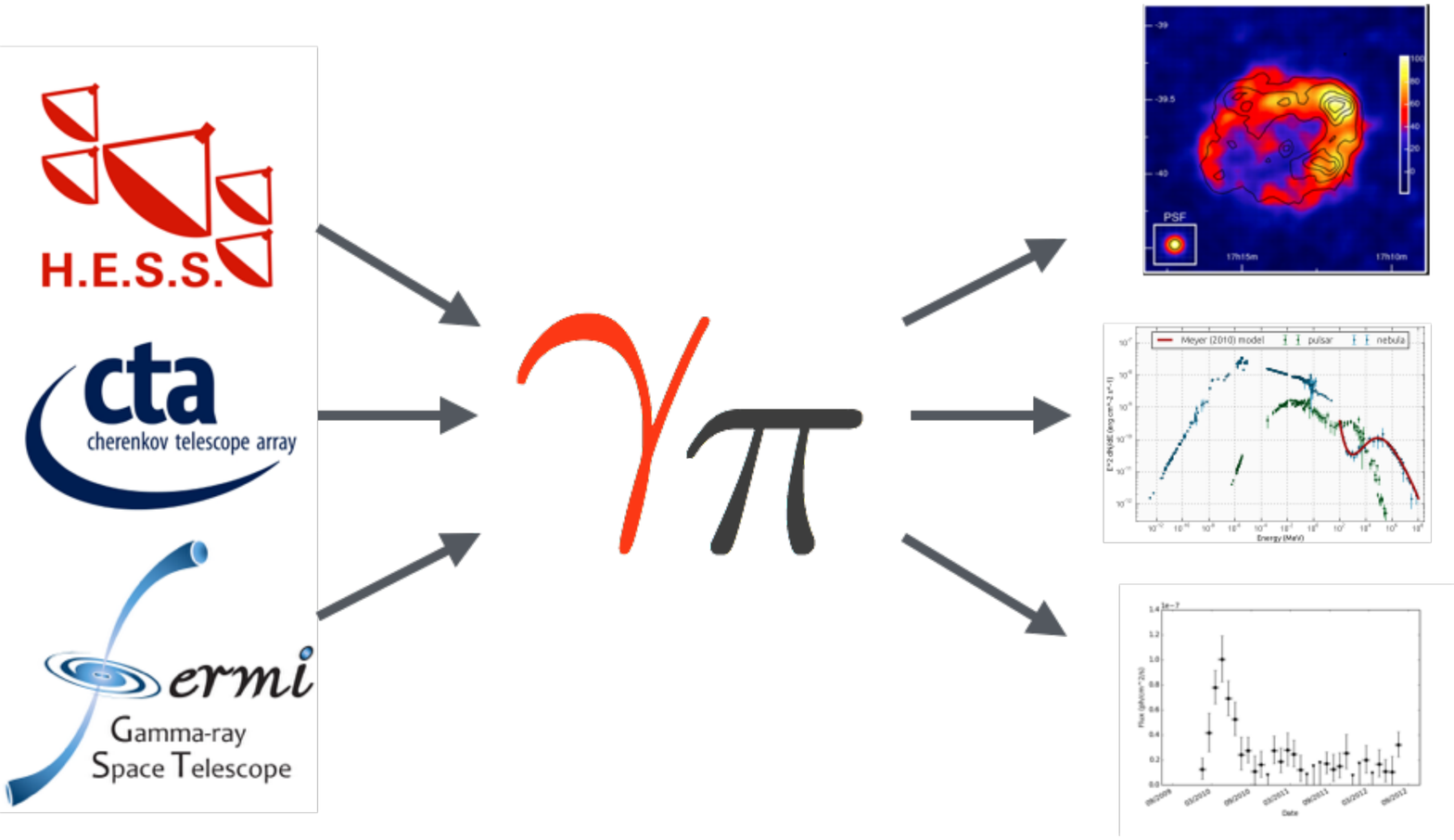}
\caption{
Gammapy is a Python package for high-level \gammaray data analysis. Using event
lists, exposures and point spread functions as input you can use it to generate
science results such as images, spectra, light curves or source catalogs.
So far it has been used to simulate and analyse H.E.S.S., CTA and \textit{Fermi}-LAT data,
hopefully it will also be applied to e.g. VERITAS, MAGIC or HAWC data in the future.
}
\label{fig:bigpictures}
\end{figure}
\subsection{What is \gammapy?}
\gammapy is an open-source Python package for \gammaray astronomy. 
Originally \gammapy started as a place to share morphology fitting Python scripts for the work on the H.E.S.S. Galactic plane survey~\cite{Deil2015} two years ago. Since that time \gammapy has grown steadily: functionality as well as development infrastructure has been improved and it has been accepted as an in-development Astropy-affiliated package. Now, in this proceeding, we would like to introduce \gammapy to the community and present our vision of \gammapy as a future community-developed, general purpose analysis toolbox for \gammaray astronomers.

The general concept of \gammapy is illustrated in Figure~\ref{fig:bigpictures}.
Based on pre-processed input data (e.g. event lists) provided by instruments such as H.E.S.S., \textit{Fermi} or CTA, \gammapy offers the high-level analysis tools to generate science results such as images,
spectra, light curves and source catalogs. By using common data structures and restriction to binned analysis techniques, all input data can be treated the same way, independent of the instrument. Research already making use of \gammapy is presented in \cite{Deil2015, Owen2015, Puehlhofer2015}.

\subsection{How to get \gammapy}
Recently \gammapy version $0.3$ was released. It is available via
the Python package index \footnote{\url{https://pypi.python.org/pypi/gammapy/}} or
using package manager tools like \href{https://pip.pypa.io/en/stable/}{pip} and
\href{http://conda.pydata.org/docs/intro.html}{conda}. \gammapy works on Linux and
Mac (Windows most likely as well, but this has not been tested yet) and is compatible with
Python 2.7 and Python 3.3 or later. Further details on requirements and installation
are available online \footnote{\url{https://gammapy.readthedocs.org/en/latest/install.html}}.
The latest documentation is available on \textit{Read the Docs}
\footnote{\url{https://gammapy.readthedocs.org/en/latest}}, including tutorials, code and analysis examples. We have also set up a mailing list for user support
and discussion \footnote{\url{https://groups.google.com/forum/\#!forum/gammapy}}.

\section{The \gammapy stack}

\begin{figure}
\centering
\includegraphics[width=0.9\textwidth]{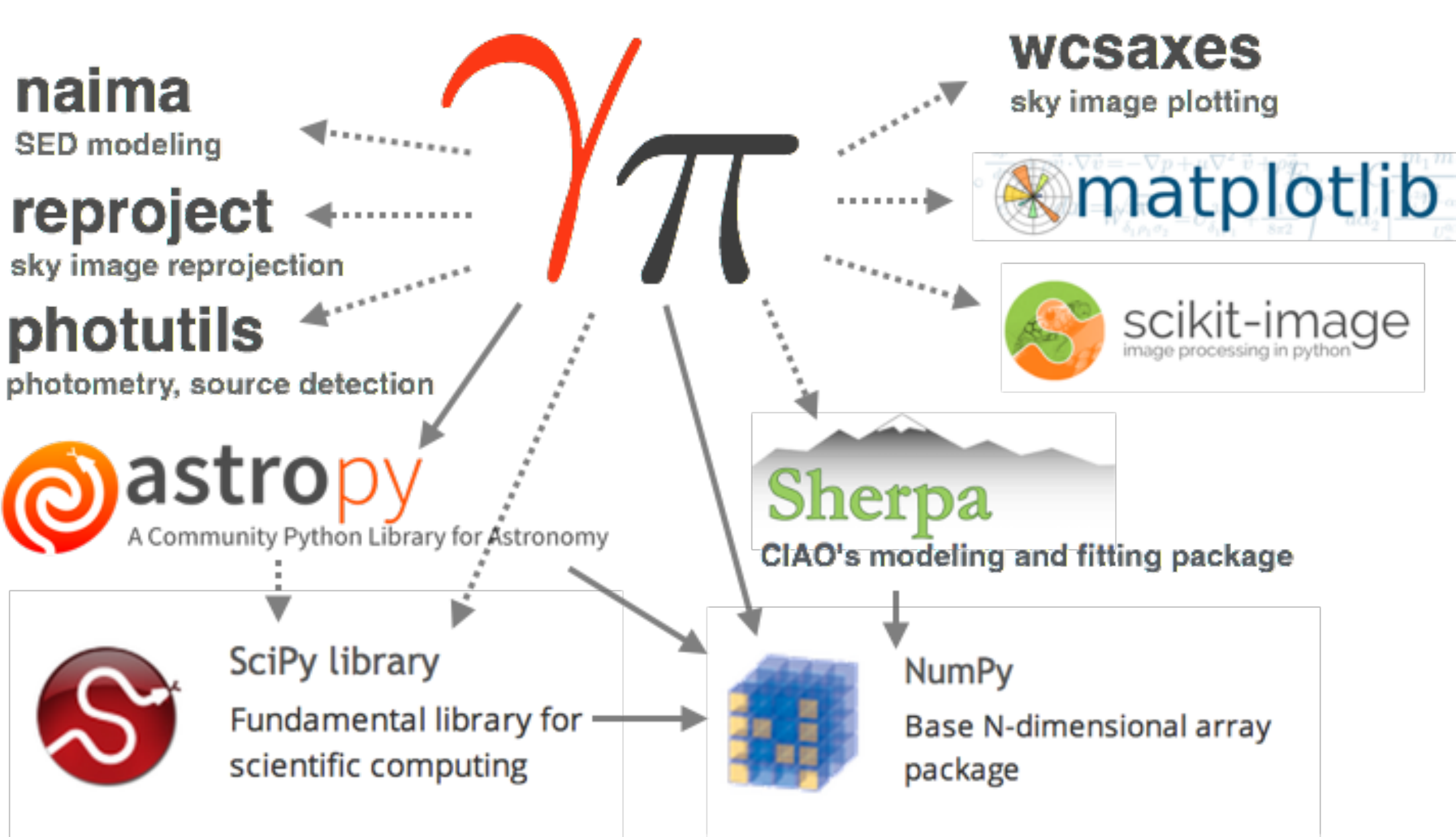}
\caption{
The Gammapy stack. Required dependencies Numpy and Astropy are illustrated with solid arrows,
optional dependencies (the rest) with dashed arrows.
}
\label{fig:dependencies}
\end{figure}

\gammapy is primarily built on the scientific Python stack. We employ Numpy and
Astropy \cite{Robitaille2013} as main dependencies and integrate other packages as optional dependencies where necessary. The current dependency
structure is illustrated in Figure~\ref{fig:dependencies}. Numpy provides the low
level data structures and the framework for numerical/array computations.
Astropy is used for higher level data structures like
tables (Table) and n-dimensional data objects (NDData), for I/O, coordinates
and WCS transformations, and handling of physical quantities with units.   Scipy
is used for advanced numerical and data processing algorithms. For
specific image processing routines we additionally use scikit-image.

To allow morphology and spectral fitting of TeV sources with \gammapy we use
the established X-ray modelling and fitting package Sherpa \cite{Refsdal2009}.
Sherpa allows interactive and scripted fitting of data sets with various spectral,
light curve and morphology models taking instrument response functions into account.
Additionally, it is possible to determine confidence levels on best-fit model parameters,
compute likelihood profiles and goodness of fit measures. Sherpa recently also became an
open-source project which makes it possible for users and external developers to contribute
missing functionality or fix issues themselves in future.

Modelling and fitting of non-thermal radiation processes to spectral energy distributions (SEDs) is provided with the
\href{https://github.com/zblz/naima}{Naima} package. It uses Markov-Chain
Monte Carlo \href{http://dan.iel.fm/emcee/}{emcee} sampling to find best-fit model
parameters of physical radiation models and thus determine the radiation mechanism
that leads to the observed emission. The radiative models available in Naima can
be called as source models in the \gammapy environment and be used, for example, to simulate
spectra for source population studies. 

Data visualizing and plotting of sky images is done with matplotlib and the
astropy-affiliated package wcsaxes. Further astropy-affilated packages we allow
as optional dependencies are photutils and reproject. Photutils is used for
source detection and photometry and reproject for re-projection of sky images. 

While this is a large number of dependencies, most of them
are optional and only needed for small specific tasks so we do not see this as a strong
disadvantage. Optional packages can be easily installed using
package managing tools like \textit{pip} or \textit{conda}. By re-using other
packages' functionality which would be complex to re-implement, new tools and techniques can be easily integrated into \gammapy to help obtain scientific results more quickly with a minimal coding effort.

\section{Development workflow}
\gammapy uses all the standard tools of modern community-driven open-source
software  development. The code repository is hosted on \textit{GitHub}
\footnote{\url{https://github.com/gammapy/gammapy}}, which allows for convenient and public
interaction of developers, contributors and users via the \textit{GitHub}
interface. This includes, for instance, \textit{feature requests} for missing
functionality, \textit{issues} for bugs or questions and \textit{pull requests}
for code contributions. The code base is continuously built and tested in
different virtual environments using the \textit{continuous integration} service
\href{https://travis-ci.org/}{Travis CI}. This helps to maintain high coding
standards and compatibility with different versions of Numpy and Astropy.

The documentation of \gammapy is generated using
\href{http://sphinx-doc.org/}{Sphinx}. An  online version of the latest
documentation is built automatically and is available on \textit{Read the Docs}
\footnote{\url{https://gammapy.readthedocs.org/en/latest}}. In addition to the main code repository we maintain
\href{https://github.com/gammapy/gammapy-extra}{gammapy-extra}. This repository
contains prepared catalog and map data and IPython notebooks with more extensive
analysis examples and tutorials.

\section{The \gammapy toolbox}
\subsection{Sub-packages}
The \gammapy code base is structured into several sub-packages where each of the packages bundle corresponding functionality in a namespace. This follows the concept of other Python packages such as Astropy and Scipy. The 
following list gives a rough overview of the different sub-packages with
a short description:

\begin{itemize}[noitemsep]
\item \href{http://gammapy.readthedocs.org/en/\gammapyversion /astro/index.html}{\lstinline$gammapy.astro$} --
Galactic population and emission models of TeV sources
\item \href{http://gammapy.readthedocs.org/en/\gammapyversion /background/index.html}{\lstinline$gammapy.background$} --
Background estimation and modeling
\item \href{http://gammapy.readthedocs.org/en/\gammapyversion /catalog/index.html}{\lstinline$gammapy.catalog$} --
\gammaray source catalog access and processing 
\item \href{http://gammapy.readthedocs.org/en/\gammapyversion /datasets/index.html}{\lstinline$gammapy.datasets$} --
Easy access to bundled and remote datasets
\item \href{http://gammapy.readthedocs.org/en/\gammapyversion /detect/index.html}{\lstinline$gammapy.detect$} --
Source detection tools and algorithms
\item \href{http://gammapy.readthedocs.org/en/\gammapyversion /detect/index.html}{\lstinline$gammapy.hspec$} --
Interface to spectral fitting with Sherpa
\item \href{http://gammapy.readthedocs.org/en/\gammapyversion /image/index.html}{\lstinline$gammapy.image$} --
Image processing and analysis tools
\item \href{http://gammapy.readthedocs.org/en/\gammapyversion /irf/index.html}{\lstinline$gammapy.irf$} --
Instrument response function (IRF) access and handling
\item \href{http://gammapy.readthedocs.org/en/\gammapyversion /morphology/index.html}{\lstinline$gammapy.morphology$} --
Morphology models and tools
\item \href{http://gammapy.readthedocs.org/en/\gammapyversion /obs/index.html}{\lstinline$gammapy.obs$} --
Observation handling
\item \href{http://gammapy.readthedocs.org/en/\gammapyversion /spectrum/index.html}{\lstinline$gammapy.spectrum$} --
Spectrum models and tools
\item \href{http://gammapy.readthedocs.org/en/\gammapyversion /stats/index.html}{\lstinline$gammapy.stats$} --
Statistical functions
\item \href{http://gammapy.readthedocs.org/en/\gammapyversion /time/index.html}{\lstinline$gammapy.time$} --
Handling of time series and \gammaray lightcurves
\item \href{http://gammapy.readthedocs.org/en/\gammapyversion /utils/index.html}{\lstinline$gammapy.utils$} --
Utility functions and classes (in sub-modules)
\end{itemize}

There are some cases (e.g. \lstinline$gammapy.spectrum.models$ and several sub-modules of\newline
\lstinline$gammapy.utils$) where the end-user functionality is exposed one level further
down in the hierarchy. This is because putting everything into the top-level
\lstinline$gammapy.spectrum$ or \lstinline$gammapy.utils$ namespace would lead to an unstructured
collection of functions and classes. A large part of \gammapy 's functionality
uses an object-oriented API. Functions are only used where it leads 
to an improved or more intuitve user interface.

\subsection{Command-line tools}
\gammapy includes various ready to use command-line tools which provide a
familiar interface to data processing for many astronomers. A complete overview of all
available tools can be found online \footnote{\url{https://gammapy.readthedocs.org/en/\gammapyversion /scripts/index.html}}.
When \gammapy is installed the user can simply type \lstinline$gammapy-$ on the command-line
and use tab completion to see the list of tools, as all \gammapy tools start
with the same \lstinline$gammapy-$ prefix. Specific information on single command-line
tools and a description of available parameters is shown when calling the corresponding
tool with the standard \lstinline$-h$ or \lstinline$--help$ option.

\section{Usage examples}
The \gammapy package should be considered as a toolbox out of which powerful
analysis scripts can be composed easily by astronomers, even if they have very
little programming experience. In the following section we present a selection
of code examples demonstrating how to set up complex analysis steps with
just a few lines of code.

\subsection{Morphology fitting command-line tool}
\gammapy includes a simple to use command-line script \lstinline$gammapy-sherpa-like$ to
perform a Poisson maximum likelihood morphology fitting using FITS files as
input. The input counts, exposure and background maps can be specified by the
corresponding parameters \lstinline$--counts$, \lstinline$--exposure$ and
\lstinline$--background$. The source model is defined in a JSON file and should
be passed using the \lstinline$--sources$ option. The model parameters for a
multi-Gaussian point spread function (PSF) can be specified in JSON format using
the \lstinline$--psf$ option.
\smallskip
\begin{lstlisting}
$ gammapy-sherpa-like --counts counts.fits --exposure exposure.fits
 --background background.fits --psf psf.json --sources sources.json
  result.json
\end{lstlisting}
Fit results and additional information are stored in \lstinline$result.json$.

\subsection{Test statistics maps}

\begin{figure}
\centering
\includegraphics[width=1.\textwidth]{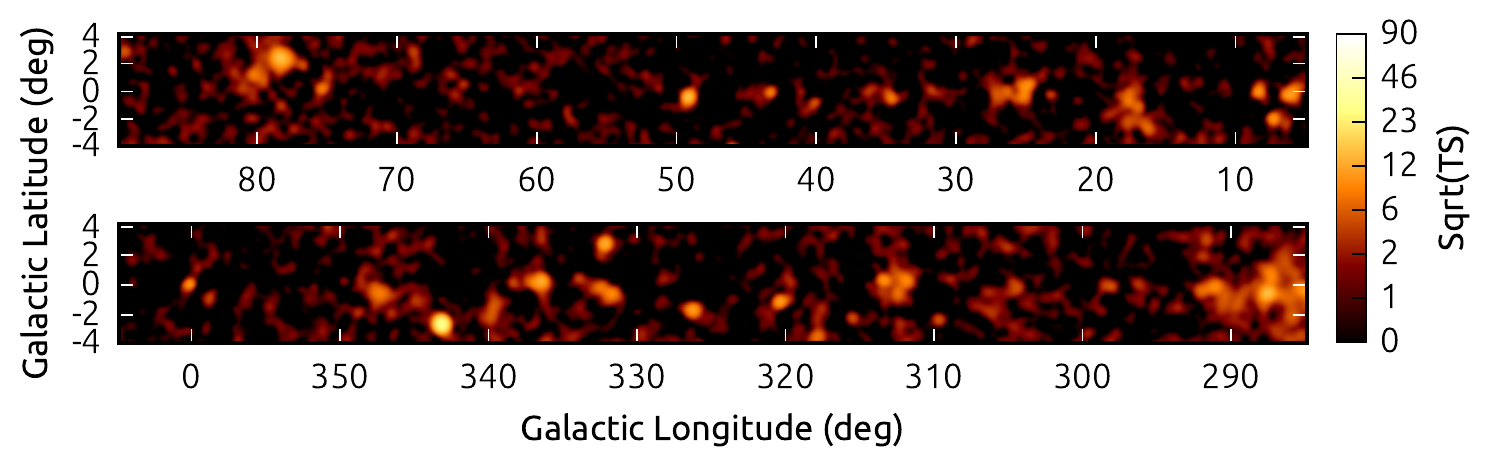}
\caption{
\textit{Fermi} survey TS map. 
}
\label{fig:fermi_ts_map}
\end{figure}

The \lstinline$gammapy.detect$
module includes a high performance \lstinline$compute_ts_map$ function to
compute test statistic (TS) maps for \gammaray survey data. The implementation
is based on the method described in \cite{Stewart2009}. As input data, the user provides counts,
background and exposure maps. The following code
example demonstrates the computation of a TS map for prepared \textit{Fermi} survey data,
which is provided in \href{https://github.com/gammapy/gammapy-extra}{gammapy-extra}. The resulting TS map is shown in Figure~\ref{fig:fermi_ts_map}.
\smallskip
\begin{lstlisting}
from astropy.io import fits
from astropy.convolution import Gaussian2DKernel
from gammapy.detect import compute_ts_map
hdu_list = fits.open('all.fits.gz')
kernel = Gaussian2DKernel(|2.5|)
result = compute_ts_map(hdu_list['On'].data,
                        hdu_list['Background'].data,
                        hdu_list['ExpGammaMap'].data, kernel)
result.write('ts_map.fits')
\end{lstlisting}

\subsection{Galactic population models}

\begin{figure}
\centering
\begin{minipage}[b]{0.48\textwidth}
%\label{fig:spiralarms}
\includegraphics[scale=1]{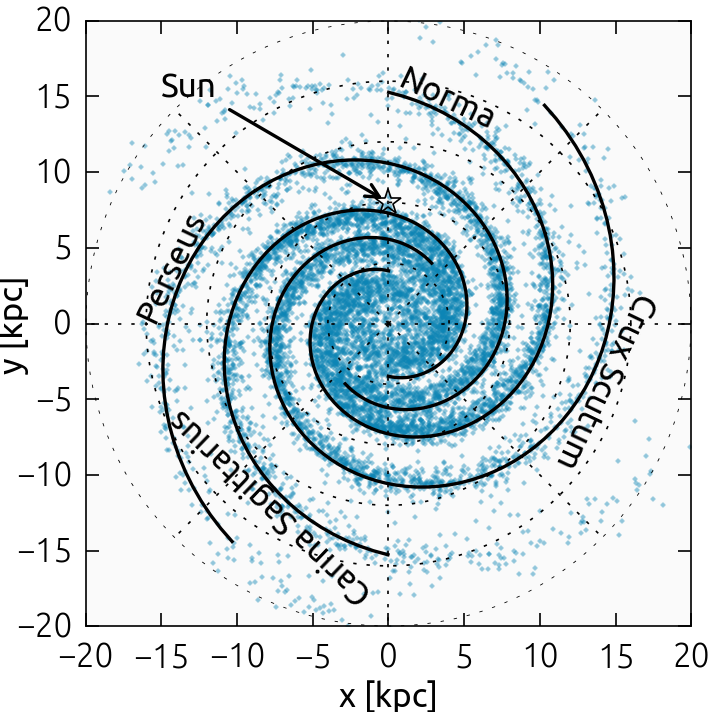}
\end{minipage}
\quad
\begin{minipage}[b]{0.48\textwidth}
%\label{fig:raddis}
\includegraphics[scale=1]{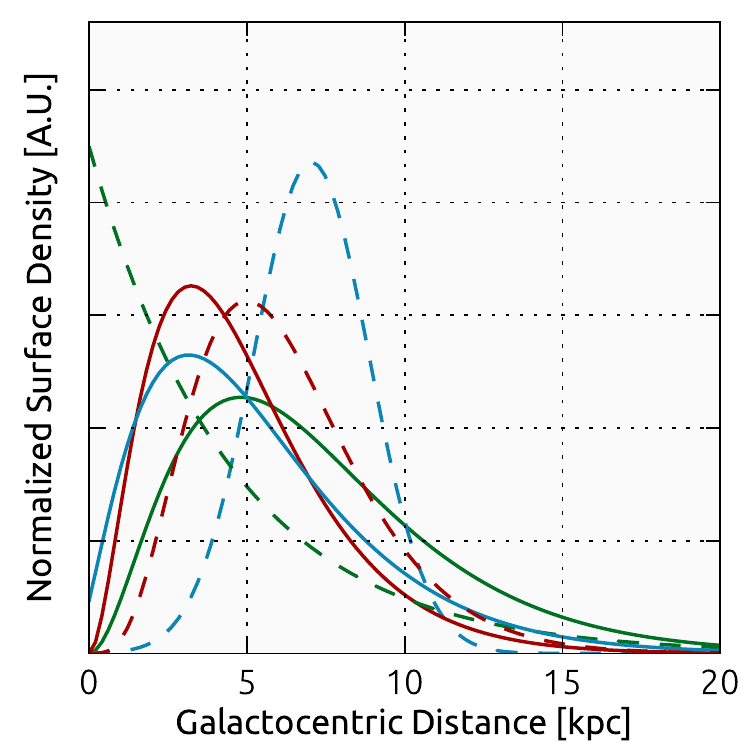}
\end{minipage}
\caption{Galactic source population simulated using \gammapy, assuming a radial
distribution of sources after \cite{Yusifov2004} and the spiral-arm model of
\cite{Vallee2008}.}
\label{fig:raddis}
\end{figure}

\gammapy also offers functionality for modeling and simulation.
The \lstinline$gammapy.astro$ sub-package provides tools to simulate Galactic
TeV source populations and source characteristics, which is useful in the
context of surveys and population studies. Figure~\ref{fig:raddis} illustrates
a source population simulated with \gammapy and the different radial distribution
models the user can choose from.

\section{Planned functionality}
\begin{figure}
\centering
\includegraphics[width=0.8\textwidth]{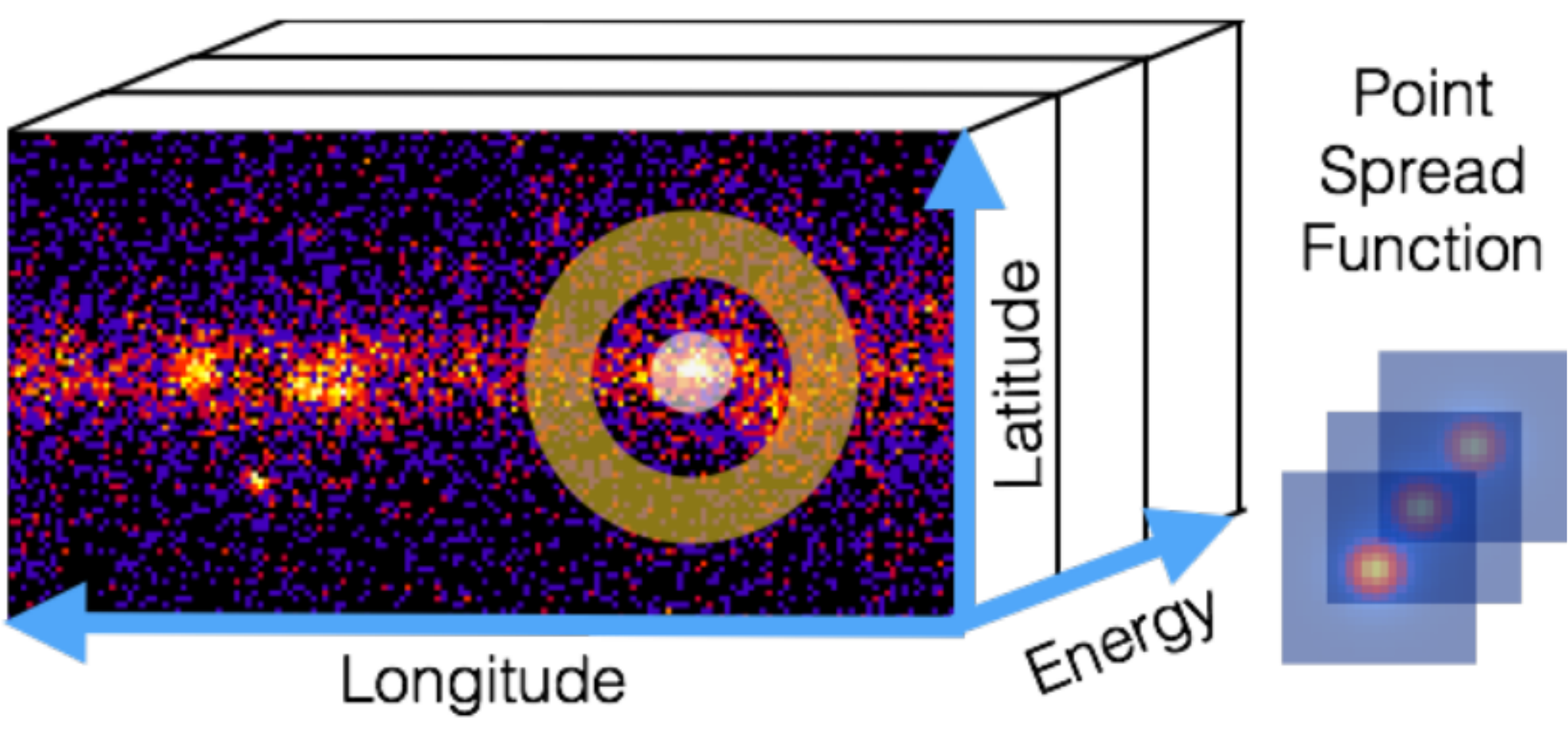}
\caption{
Gammapy data model illustration. Binned analysis of lon-lat-energy cube data is
supported via joint likelihood analysis of one image per energy bin.
On-off-region based spectral analysis is supported as well.
}
\label{fig:data-model}
\end{figure}

\gammapy already includes key analysis features like morphology and spectrum fitting but both are currently limited to either 2D image-based data or 1D
spectral data. As a major next step we plan to support joint likelihood fitting of datasets, as illustrated in Figure~\ref{fig:data-model}. Events are binned into longitude, latitude and energy cubes and fitted simultaneously with spectral and spatial models taking energy-dependent background, exposure and point spread function (PSF) into account.

 This approach will allow joint likelihood analysis across different experiments. For instance simultaneous likelihood fitting of Fermi and H.E.S.S.
data. 

 Support for un-binned analyses is not planned.

\section{Summary}
\gammapy 0.3 is still alpha quality software. It is a package where standard \gammaray analyses are available on the one hand while, on the other hand, integration and prototyping of new methods is easily possible. So far it only contains limited
functionality but the setup of documentation, testing and deployment is already
very advanced. It's scope will continously grow and we hope that many users and developers show interest in open and reproducible \gammaray astronomy with Python.
As long-term goal we would like \gammapy to turn into a fully community-developed package. So all contributions to \gammapy are welcome! If you don't know how to turn your scripts into production-quality, reusable code, please get in touch with us (e.g. using the mailing list) and we will help you get there!

For further information on how \gammapy and other tools are being used for \hess
data analysis, we encourage you to look at \cite{DeilHost2015}.

\bibliography{gammapy-icrc2015}

\begin{thebibliography}{1}
\bibitem{Puehlhofer2015}
G.~{P\"uhlhofer et al., for the H.E.S.S. collaboration}, {\it {Search for new
  supernova remnant shells in the Galactic plane with {H.E.S.S.}}},  in {\em
  \textit{these proceedings}}, 2015.
\end{thebibliography}
\bibliographystyle{JHEP}

\end{document}